\documentclass[aps,prb,reprint,noeprint,superscriptaddress,nobibnotes,longbibliography]{revtex4-2}
\pdfoutput=1
\usepackage[utf8]{inputenc}
\usepackage[english]{babel}
\usepackage{microtype}
\usepackage{amsmath,amssymb,amsfonts}
\usepackage{braket}
\usepackage{bm}
\usepackage{graphicx}
\usepackage{booktabs}

\usepackage[colorlinks=true,unicode=true,pdfborder={0 0 0},allcolors=blue
]{hyperref}

\renewcommand{\vec}[1]{\bm{#1}}
\newcommand{\im}{\mathrm{i}}
\newcommand{\e}{\mathrm{e}}

\newcommand{\vk}{{\vec k}}
\newcommand{\vq}{{\vec q}}

\newcommand{\VVT}[2]{\begin{pmatrix}#1, & #2\end{pmatrix}}

\newcommand{\VVVT}[3]{\begin{pmatrix}#1, & #2, & #3\end{pmatrix}}

\begin{document}

\title{Current-assisted Raman activation of the Higgs mode in superconductors}

\author{Matteo Puviani}
\affiliation{Max Planck Institute for Solid State Research,
70569 Stuttgart, Germany}

\author{Lukas Schwarz}
\affiliation{Max Planck Institute for Solid State Research,
70569 Stuttgart, Germany}

\author{Xiao-Xiao Zhang}
\affiliation{Department of Physics and Astronomy \& Stewart Blusson Quantum Matter Institute, University of British Columbia, Vancouver, Canada BC V6T 1Z4}

\author{Stefan Kaiser}
\affiliation{Max Planck Institute for Solid State Research,
70569 Stuttgart, Germany}
\affiliation{4th Physics Institute and Research Center SCoPE,
University of Stuttgart, 70569 Stuttgart, Germany}

\author{Dirk Manske}
\email{d.manske@fkf.mpg.de}
\affiliation{Max Planck Institute for Solid State Research,
70569 Stuttgart, Germany}

\date{\today}

\begin{abstract}
The Higgs mode in superconductors is a scalar mode
without electric or magnetic dipole moment.
Thus, it is commonly believed that its excitation is restricted
to a nonlinear two-photon Raman process.
However, recent efforts have shown
that a linear excitation in the presence of a supercurrent is possible,
resulting in a new resonant enhancement at $\Omega=2\Delta$
with the driving light frequency $\Omega$
and the energy of the Higgs mode $2\Delta$.
This is in contrast to the usual $2\Omega = 2\Delta$ resonance condition
found in nonlinear third-harmonic generation experiments.
In this communication, we show
that such a linear excitation can still be described
as an effective Raman two-photon process,
with one photon at $\omega=2\Delta$ and one virtual photon at $\omega=0$
which represents the dc supercurrent.
At the same time we demonstrate that a straightforward infrared activation
with a single photon excitation is negligible.
Moreover, we give a general context to our theory, providing an explanation
for how the excitation of the Higgs mode in both
THz quench and drive experiments can be understood
within a conventional difference-frequency generation
or sum-frequency generation process, respectively.
In such a picture, the observed new resonance condition $\Omega = 2\Delta$
is just a special case.
With the same approach,
we further discuss another recent experiment,
where we find a suppression of odd order higher harmonics
in the presence of a dc supercurrent.
\end{abstract}

\maketitle

\textit{Introduction.}
Light excitation of collective modes in condensed matter physics
is typically realized due to an infrared or Raman coupling
of light to the system.
This corresponds to a one- or two-photon process,
i.e. a linear or a nonlinear coupling.
If the mode does not have a dipole moment,
a linear activation is forbidden,
such that the only allowed process is due to the nonlinear Raman effect.
This is true for the Higgs mode in superconductors,
which is a collective oscillation of the amplitude of the order parameter
\cite{JLowTempPhys.126.901,annurev.varma2015}.
Its observation so far was realized either in a quench-probe
\cite{PhysRevLett.111.057002}
or a periodic driven setup
\cite{Science.345.1145,PhysRevB.96.020505,NatCommun.11.1793},
where in both cases the coupling of light to the superconductor
can be described by a quadratic nonlinear effective Raman process.
Due to this nonlinear coupling,
light of frequency $\Omega$ drives the system effectively
with a frequency of $2\Omega$
leading to enforced $2\Omega$ oscillations of the order parameter.
A tuning of the effective driving frequency
to the energy of the Higgs mode at $2\Delta$
leads to an enhancement of the oscillations
and a resonance peak in the spectrum at $2\Omega = 2\Delta$.
This second-harmonic component in the order parameter oscillation
translates into third-harmonic generation (THG) in emission
and the resonance is observable in the emitted THG signal
\cite{PhysRevB.92.064508,PhysRevB.93.180507}.

Recent studies have shown
that a linear coupling of light to the condensate is possible
in the presence of a supercurrent \cite{PhysRevLett.118.047001}.
Hereby, the current provides a momentum
to the center of mass of the condensate,
such that the scalar Higgs mode can be excited with a single photon.
With this, the equilibrium forbidden linear driving
with the frequency $\Omega$ is now possible
due to the breaking of equilibrium inversion symmetry
and enforces the order parameter to oscillate at just $\Omega$.
As a result, the first-harmonic oscillation of the order parameter shows
a new resonance condition at $\Omega = 2\Delta$.
In an emission experiment,
second-harmonic generation (SHG) would occur
and the new resonance is then observable in the emitted SHG signal.
Furthermore, the effect is also visible in the linear response
and indeed, an enhancement in the optical conductivity
was observed in the superconductor NbN,
which was driven by a dc supercurrent \cite{PhysRevLett.122.257001}.

In this communication we extend previous theoretical studies of
current- and light-driven superconductors,
providing a new interpretation of the experimental findings.
We argue that this coupling should not be understood as an infrared activation
of the Higgs mode in a single photon process
but rather as a current-assisted Raman interaction,
where the excitation is still realized
with a two-photon coupling to the condensate.
In the present work, we exploit this effect to interpret experimental results
and to predict further experimental outcomes.
In this two-photon process, one photon is at the energy $\omega = \Omega$
and the other photon is a virtual photon
with $\omega = 0$ given by the dc supercurrent.
We find that a single photon infrared activation is negligible
compared to such an effective Raman process.

Distinguishing the current-assisted Higgs excitation
as being either infrared or Raman-like is fundamental
and has important implications for designing future experiments
to excite and investigate the Higgs mode
in the flourishing field of Higgs spectroscopy.
As there is an ongoing debate on how this activation can be understood
\cite{PhysRevLett.122.257001,NatPhoton.10.707,NatPhys.15.341},
this communication suggests a consistent interpretation,
which was not given in the previous theoretical study
\cite{PhysRevLett.118.047001}.
Furthermore, the description of the excitation process is more general
and allows to describe the usual nonlinear two-photon activation
of the Higgs mode, or more generally any collective mode,
in quench or drive experiments
with a sum-frequency generation (SFG)
or difference-frequency generation (DFG) scheme.
Within this picture,
the current-assisted excitation just represents a special case
with one photon energy set to zero.

\textit{Current-assisted Raman.}
\label{sec:car}
To describe the coupling of light to current-carrying superconductors,
we start with a phenomenological description
using a Lagrangian of Ginzburg-Landau type.
For the complex superconducting order parameter $\psi(\vec r, t)$
coupled to a gauge field $A_\mu = (\phi, \vec{A}(t))$
where we choose $\phi = 0$,
the time- and space-dependent Lagrangian reads
\begin{align}
    \mathcal L &= (D_\mu \psi)^* (D^\mu \psi)
        - V(\psi) - \frac 14 F_{\mu \nu} F^{\mu \nu}\,,
\end{align}
where the potential $V(\psi) = \alpha |\psi|^2 + \frac \beta 2 |\psi|^4$
has the shape of a Mexican hat in the superconducting state with $\alpha < 0$.
The gauge covariant derivative reads
$D_\mu = \partial_\mu + \im e A_\mu$
with the effective charge $e$ and the electromagnetic field tensor
$F_{\mu \nu} = \partial_\mu A_\nu - \partial_\nu A_\mu$.
We are interested in small fluctuations around the groundstate value
$|\psi_0| = \sqrt{-\alpha/\beta}$,
thus, we use an ansatz
$\psi(\vec r, t) = (\psi_0 + H(\vec r,t))\e^{\im\theta(\vec r, t)}$
to describe amplitude (Higgs) fluctuations $H$
and phase (Goldstone) fluctuations $\theta$.
Neglecting higher orders and constant terms,
the resulting Lagrangian for the fluctuations reads
\begin{align}
    \mathcal L &= \left(\partial_\mu H
        - \im e\left(A_\mu + \frac 1 e \partial_\mu\theta \right)(\psi_0 + H)
        \right)\notag\\
        &\quad \times \left(\partial^\mu H
        + \im e\left(A^\mu + \frac 1 e \partial^\mu \theta\right)(\psi_0 + H)
        \right) \notag\\
        &\quad +2\alpha H^2 - \frac 1 4 F_{\mu\nu}F^{\mu\nu}\,.
\end{align}
We now consider the situation
in which the superconductor is at the same time
driven by a homogeneous light field with frequency $\Omega$
and in which a dc supercurrent is injected.
Thus, we write the vector potential as
$\vec{A}(t) = \vec A_0 \e^{\im\Omega t} + \vec Q/e$,
where the condensate momentum $\vec Q$ is defined
by the gauge invariant current
$\vec j = e n_s \hbar/m (e\vec A - \nabla \theta) = e n_s \hbar/m \vec{Q}$,
with $n_s$ being the superconducting density,
$m$ and $e$ the electron effective mass and charge, respectively
\cite{arxiv.1908.10879}.
Due to the Anderson-Higgs mechanism,
the phase fluctuations can be gauged out by a choice of $\chi = -\theta$,
where $\psi' = \psi \e^{\im \chi}$
and the redefined vector potential
$A_\mu' = A_\mu - \frac 1 e \partial_\mu \chi$ \cite{PhysRev.130.439}.
Dropping the primes, the Lagrangian then reads
\begin{align}
    \mathcal L & = (\partial_\mu H)(\partial^\mu H)
        + 2\alpha H^2 - \frac 1 4 F_{\mu\nu}F^{\mu\nu}\notag\\
    &\quad + e^2 \psi_0^2 A_\mu A^\mu + 2e^2\psi_0 A_\mu A^\mu H\,.
\end{align}
The term $\mathcal L_{\mathrm{int}} = 2e^2\psi_0 A_\mu A^\mu H$
describes the coupling between the gauge field and Higgs and reads explicitly
\begin{align}
    \mathcal L_{\mathrm{int}}
        = 2e^2\psi_0(\vec A^2 H + \frac{2}{e} \vec{Q} \vec A H
            + \frac{1}{e^2} \vec{Q}^2 H)\,.
\end{align}
We observe that for finite $\vec Q$ a linear coupling term arises.
The equation of motion for the amplitude fluctuations $H$ at $q=0$ reads
\begin{align}
    \partial_t^2 H &= 2\alpha H + e^2\psi_0 \vec A^2
        + 2e \psi_0 \vec{Q} \vec A
        + \psi_0 \vec{Q}^2
\end{align}
and the stationary solution is easily obtained as
\begin{align}
    H = \frac{\psi_0 Q^2}{\omega_\mathrm H^2}
    - \frac{2 e \psi_0 Q A_0}{\Omega^2 - \omega_\mathrm H^2}\e^{\im\Omega t}
    - \frac{e^2\psi_0A_0^2}{4\Omega^2 - \omega_\mathrm H^2}\e^{2\im\Omega t}\,,
\end{align}
where the energy of the Higgs mode $\omega_\mathrm H^2 = -2\alpha$
is obtained for $A_\mu=0$.
From the microscopic theory it is known that $\omega_\mathrm H = 2\Delta$
\cite{JLowTempPhys.126.901}.
For $\vec Q = 0$, only the nonlinear driven oscillations exists
with $\omega = 2\Omega$ resonating at $2\Omega = 2\Delta$.
For $\vec Q \neq 0$, a linear coupling is possible resulting in an oscillation
with $\omega = \Omega$ resonating at $\Omega = 2\Delta$.
Using similar arguments, this result of linear coupling was first derived
in \cite{PhysRevLett.118.047001}.
In general, one could have included an additional
nonrelativistic Gross-Pitaevskii-like term $\propto  \psi^*D_0\psi$
in addition to the relativistic Klein-Gordon-like description of the dynamics.
However, it is known that in the nonrelativistic case,
there is no distinct Higgs mode
as amplitude and phase channel are coupled \cite{annurev.varma2015}.
Despite the fact that a current would support a particle-hole breaking term,
its contribution cannot be large.
Otherwise, a resonance at $\omega_\mathrm H$ is not explainable.
\begin{figure}[t]
    \centering
    \includegraphics[width=\columnwidth]{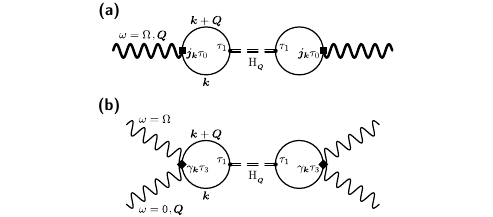}
    \caption{\label{fig:feynman}%
    Feynman diagrams describing excitation of Higgs mode
    in current-carrying state with condensate momentum $\vec Q$
    and light frequency $\Omega$.
    a) Single-photon infrared excitation
    with first-order derivative vertex interaction
    $j_\vk \propto \partial_\vk \epsilon_\vk$.
    b) Effective Raman excitation
    with second-order derivative vertex interaction
    $\gamma_\vk \propto \partial^2_\vk \epsilon_\vk$.
    Wiggly, wiggly bold, solid and double dashed lines represent
    photon, photon interacting with current, electron and Higgs propagator.}
\end{figure}%

With the most general Lagrangian approach
we have demonstrated that the coupling of light to the Higgs mode
in a superconductor can be linear in the presence of a supercurrent.
A normal dissipative current would not be sufficient,
as a nonzero superfluid momentum has to be present
to break time-reversal symmetry.
In order for this effect to be induced by a normal current,
electrons would have to interact at the same time with the light
and the Cooper pairs of the condensate.
A minimal scheme representing the interaction of one photon
with the moving condensate suggested by the Lagrangian description
is the infrared activation of the Higgs mode,
depicted in Fig.~\ref{fig:feynman}(a).
Thus, we calculate the current-current correlation function
in the presence of the Higgs mode
using this diagram.
The amplitude of the transmitted electric field is given by
$E(\vec Q, \Omega) \propto %
\vec Q\vec A |\chi^\mathrm H_{\mathrm j\mathrm j}(\vec Q, \Omega)|$
\cite{PhysRevB.93.180507,RevModPhys.79.175,PhysRevB.100.165131}.
The susceptibility $\chi^H_{\mathrm j\mathrm j}(\vec q, \Omega)$ is given by
\begin{align}
    \chi_{\mathrm j\mathrm j}^H (\vq, \Omega)
        &= H(\vq, \Omega) \chi_{\mathrm j1}^2 (\vq, \Omega)
        = - \frac{\chi_{\mathrm j1}^2 (\vq, \Omega)}
            {2/V + \chi_{11} (\vq, \Omega)}
\end{align}
with the condensate pairing interaction $V$
and the Higgs propagator
$H(\vq, \Omega) = - (2/V + \chi_{11}(\vq, \Omega))^{-1}$.
We use the indices of the susceptibility
to denote the vertices of the corresponding bubbles:
$j$ for the current channel $j_k \tau_0$
and $1$ for the amplitude channel $\tau_1$.
The vertices are written as matrices in Nambu space
\cite{PhysRev.117.648}
and $\tau_i$ are the Pauli matrices.
The susceptibility $\chi_{\mathrm j1} (\vq, \Omega)$ reads
after analytic continuation $\im\omega_n \rightarrow \Omega$ of the expression
\begin{align}
    & \chi_{\mathrm j1} (\vq, \im\omega_n) = \sum_\vk j_\vk f_\vk
        \iint \mathrm d \omega_1\, \mathrm d \omega_2\,
        \frac{\Delta_\vk \omega_2 + \Delta_{\vk+\vq} \omega_1}
        {2 E_\vk E_{\vk+\vq}} \notag\\
        &\qquad
        \times \frac{n_\mathrm F (\omega_1) - n_\mathrm F(\omega_2)}
        {\omega_1 - \omega_2 + i \omega_n} \left[
            \delta(\omega_1 - E_\vk) - \delta(\omega_1 + E_\vk )
        \right] \notag\\
        &\qquad \times \left[
            \delta(\omega_2 - E_{\vk+\vq}) - \delta(\omega_2 + E_{\vk+\vq} )
        \right] \,,
\end{align}
with $j_\vk = \partial_\vk \epsilon_\vk$,
electron dispersion $\epsilon_\vk$,
energy gap $\Delta_\vk$,
gap symmetry $f_\vk$,
quasiparticle energy $E_\vk = \sqrt{|\Delta_\vk^2| + \epsilon_\vk^2}$
and Fermi function $n_\mathrm F$.
Details of the calculation
can be found in the supplemental material \cite{supplement}.
In the limit $\vq \rightarrow 0$, the term $\Delta_\vk (\omega_1 + \omega_2)$
is nonzero only when $\omega_1 = \omega_2 = E_\vk$,
but in this case,
$n_\mathrm F(E_\vk) - n_\mathrm F (E_\vk) = 0$,
thus, $\chi_{\mathrm j1} (\vq = 0, \Omega) = 0$ is identically zero.
For finite but small $\vq = \vec Q$,
the contribution of this diagram is still small.
For very large $\vec Q$,
the contribution gets comparable to the Raman process,
which we will discuss next.
However, it is no longer peaked at $\Omega = 2\Delta$,
but is shifted to higher energies
reflecting the quadratic dispersion of the Higgs mode
\cite{PhysRevB.92.064508}.
This is shown in more detail in the supplemental material \cite{supplement}.
As no such shift is observed experimentally,
the value of $\vec Q$ must be small.
Using parameters from the experiment in \cite{PhysRevLett.122.257001},
we estimate the value of the supercurrent induced momentum
in units of the lattice constant $1/a_0 =10^8$\,cm$^{-1}$ as
$Qa_0 = \frac{m}{e n_\mathrm s \hbar} j a_0 \approx 3.7 \cdot 10^{-4}$
with electron mass $m=m_\mathrm e$ and charge $e$,
superfluid density $n_\mathrm s = 5.4\cdot 10^{20}$\,cm$^{-3}$
and current density $j = 3.7 \cdot 10^{6}$\,A/cm$^2$.
Thus, the contribution from the infrared diagram
in the current-carrying state is negligible.

Since such an infrared linear coupling has difficulties
to explain the resonance at $\Omega = 2\Delta$,
we considered a microscopic description using BCS theory
with inversion symmetry breaking due to the supercurrent.
The BCS Hamiltonian within Anderson pseudospin formulation
\cite{PhysRev.112.1900} can be written as
$H = \sum_\vk \vec b_\vk \vec \sigma_\vk$
with the definition of the pseudospin
$\vec \sigma_\vk = \frac 1 2 \Psi_\vk^\dagger \vec \tau \Psi_\vk$,
where $\Psi_\vk^\dagger = \VVT{c_{\vk\uparrow}^\dagger}{c_{-\vk\downarrow}}$
is the Nambu-Gorkov spinor
and $\vec \tau$ the vector of Pauli matrices.
The pseudomagnetic field $\vec b_\vk$
for a real gap $\Delta_\vk = \Delta f_\vk$,
driving amplitude $\vec A(t) = \vec A_0\sin(\Omega t)$
and current induced momentum $\vec Q$ reads
$\vec b_\vk^\top(t) = \VVVT{-2\Delta_\vk(t)}{0}%
{\epsilon_{\vk-e\vec A(t)-\vec Q} + \epsilon_{\vk+e\vec A(t)+\vec Q}}$.
Neglecting the imaginary part of the gap,
which is unimportant for our discussion,
the gap equation expressed in the pseudospin picture reads
$\Delta(t) = V \sum_\vk f_\vk \braket{\sigma_\vk^x}(t)$
where the function $f_\vk$ describes the symmetry of the gap
and $V$ is the pairing interaction.
We expand the $z$-component of the pseudomagnetic field
in powers of $\vec A$ and obtain
\begin{align}
    b_\vk^z &\approx 2\epsilon_{\vk}
    + \sum_{ij} \partial_{ij}^2\epsilon_\vk
    \Big(e^2A_i A_j + 2e A_iQ_j + Q_iQ_j\Big)\,,
    \label{eq:eps_expansion}
\end{align}
where $\partial^2_{ij}$ is the partial derivative
with respect to $k_i$ and $k_j$.
\begin{figure}[t]
    \centering
    \includegraphics[width=\columnwidth]{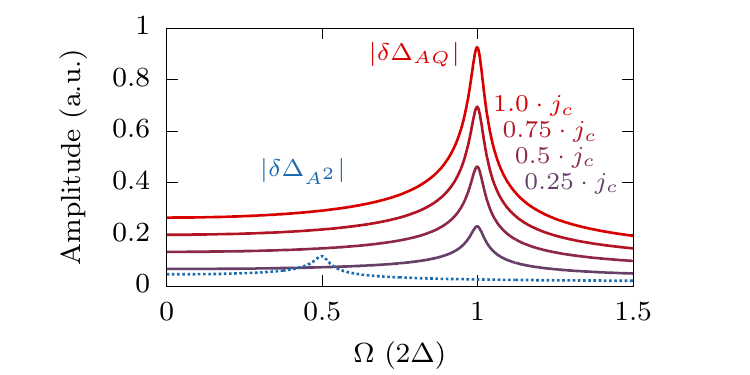}
    \caption{\label{fig:resonance}%
    Amplitudes of the induced gap oscillations from Eq.~\eqref{eq:deltat}
    for $s$-wave superconductor
    in units of $\omega_\mathrm H = 2\Delta$ with $f(\varphi) =1$
    and varying current strength using values of the experiment
    \cite{PhysRevLett.122.257001}
    (see suplemental material for details and numerical values).
    Resonances of the driving light with the Higgs mode appear
    at $2\Omega = 2\Delta$ without (blue)
    and at $\Omega = 2\Delta$ (red) with supercurrent.}
\end{figure}%
Hereby, the first order terms $\pm \sum_i \partial_i \varepsilon_{\vec k} (e A_i + Q_i)$ cancel due to parity,
corresponding to the vanishing of the infrared diagram in  Fig.~\ref{fig:feynman}(a)
for $\vec q \rightarrow 0$.
However, as for the Lagrangian description developed before,
for finite $\vec Q$, a new term linear in $\vec A$ arises.
With this approach we see from the expansion in Eq.~\eqref{eq:eps_expansion}
that both the linear and quadratic couplings are proportional
to the second derivative of the band dispersion,
giving rise to the Raman vertex coefficient $\gamma_{\vec k}$
in the effective mass approximation \cite{PhysRevLett.72.396}.
In addition, the coupling occurs in the $z$-component
of the pseudomagnetic field,
i.e. the $\tau_3$ channel, reflecting a Raman coupling
\cite{RevModPhys.79.175,PhysRevB.65.144304}.
Thus, the pseudospin description contains more information
than the Lagrangian formulation,
and no assumptions on the nature of the coupling has been made.

We now proceed by evaluating the linearized Bloch equations
$\partial_t \vec \sigma_\vk = \vec b_\vk \times \vec \sigma_\vk$
for small deviations $\delta\Delta(t)$ from the equilibrium value $\Delta$,
from which the observable higher-order current
$j^H(t) \propto \delta\Delta(t)A(t)$
can be obtained \cite{PhysRevB.92.064508}.
The solution reads
$\delta\Delta(t) = \delta\Delta_{A^2}(t) + \delta\Delta_{AQ}(t)
    + \delta\Delta_{Q^2}(t)$
where each term arises from one of the expression
$\propto A_iA_j$, $\propto A_iQ_j$ and $\propto Q_iQ_j$
in Eq.~\eqref{eq:eps_expansion}.
Assuming $\epsilon_\vk = \epsilon(|\vk|)$ and $f_\vk = f(\varphi)$
depending only on the polar angle $\varphi$
and neglecting the trivial term $\delta\Delta_{Q^2}$,
the remaining two terms read explicitly in the long-time limit
\begin{subequations}
\begin{align}
    \delta \Delta_{A^2}(t) & \propto
        \frac{e^2A_0^2\Omega \cos{(2\Omega t)}}
        {\int \text{d}\varphi f^2 \sqrt{\Delta^2 f^2 - \Omega^2}
        \sin^{-1}{\left(\frac{\Omega}{\Delta |f|}\right)}}\,,\\
    \delta \Delta_{AQ}(t) &\propto
        \frac{4eA_0Q\Omega \sin{(\Omega t)}}
        {\int \text{d}\varphi f^2 \sqrt{4\Delta^2 f^2 - \Omega^2}
        \sin^{-1}{\left(\frac{\Omega}{2\Delta |f|}\right)}}\,.
\end{align}%
\label{eq:deltat}%
\end{subequations}
The first term $\delta \Delta_{A^2}(t)$
describes the $2\Omega$ oscillations of the order parameter
resonating at $2\Omega = 2\Delta$
which are induced by the usual quadratic coupling \cite{PhysRevB.92.064508}.
The second term $\delta \Delta_{QA}(t)$
describes the equilibrium forbidden $\Omega$ oscillations
of the order parameter with a new resonance at $\Omega = 2\Delta$,
which is only present for finite condensate momentum $Q$.
The amplitudes of both terms are shown exemplary
in Fig.~\ref{fig:resonance} for $s$-wave symmetry with $f(\varphi) = 1$
and different strengths for the current
using values of the experiment \cite{PhysRevLett.122.257001}
(see supplemental material \cite{supplement} for parameter details
and an evaluation for $d$-wave).
One observes that the resonance peak in the current-activated term
is comparable in size or even exceeds the usual quadratic term
in the range of experimental reachable values.

In a diagrammatic representation,
we might describe both processes,
the quadratic and linear coupling,
with an effective Raman vertex as shown in Fig.~\ref{fig:feynman}(b).
For the quadratic coupling,
each incident photon line corresponds to one photon
of frequency $\omega = \Omega$,
resulting in the $2\Omega = 2\Delta$ resonance.
For the linear coupling,
one incident photon corresponds to the light photon with $\omega = \Omega$,
whereas the second incident photon line is a virtual photon at $\omega = 0$,
representing the dc supercurrent.
An evaluation of the diagram shows
that it is equivalent to the linearized solution in the pseudospin formalism
(see supplemental material \cite{supplement} for details).
This mechanism also has an analogy to the Higgs excitation
in the superfluid phase of ultracold bosons \cite{PhysRevB.75.085106}.
Here, a nonzero expectation value of a boson operator
due to the superfluid condensate enables a linear coupling
to the vector potential
similar to the finite $Q$ of the supercurrent.
Thus, such a current-assisted Raman diagram
can explain the $\Omega = 2\Delta$ resonance in the current-carrying state.

\begin{figure}[t]
    \centering
    \includegraphics[width=\columnwidth]{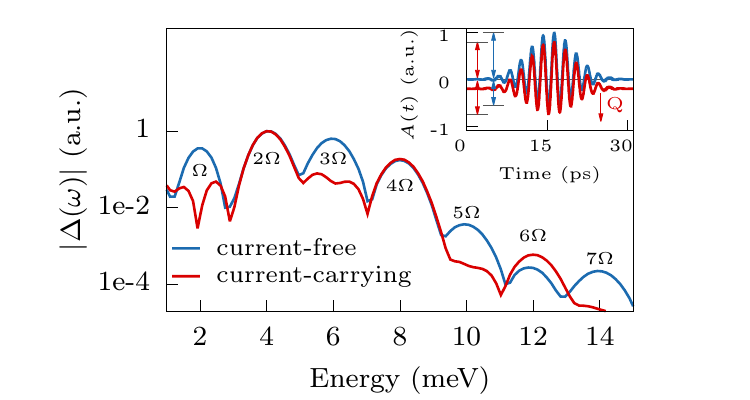}
    \caption{\label{fig:asym}%
    Spectrum of gap oscillations induced by an asymmetrically shaped
    vector potential as observed in \cite{NatPhoton.10.707} (blue).
    Note that in the current-free state,
    the positive amplitude is larger than the negative amplitude
    (as indicated by the blue arrows), while in the current-carrying state,
    the symmetry is partly restored (red arrows).
    The asymmetric pulse dynamically induces a dc supercurrent component which
    allows even and odd order higher harmonics.
    If an external supercurrent with momentum $Q$
    is applied with opposite direction,
    the odd order higher harmonics get suppressed (red curve).
    Details about the calculation can be found
    in the supplemental material \cite{supplement}.
    }
\end{figure}

Base on this interpretation,
we can also understand a recent experiment \cite{NatPhoton.10.707},
where a strong THz field dynamically induces a dc component
due to an effective asymmetric pulse shape resulting from nonlinear effects.
The situation can be described in analogy to the here discussed case
by a current-assisted Raman activation.
To make this explicit,
we model this experiment in accordance to \cite{NatPhoton.10.707}
using an asymmetrically shaped vector potential
$A(t) \propto A_0 (\sin(\Omega t)+\kappa)/(1+\kappa)$,
where $\kappa \neq 0$ determines the asymmetry,
and solve the Bloch equations numerically
(see supplemental material \cite{supplement} for details).
The vector potential and the resulting even and odd order higher harmonics
in the gap oscillations can be seen in Fig.~\ref{fig:asym} (blue curves).
In order to demonstrate the equivalence
between the dynamically induced dc component by the asymmetric vector potential
and the external dc current,
we consider the situation where both effects are included.
If an additional external supercurrent with momentum $Q$
is applied in opposite direction to the dynamically driven supercurrent,
i.e. $A(t) + Q/e$ with $Q < 0$,
the inversion symmetry can be partially restored,
giving rise to a suppression of the odd order higher harmonics (red curves).
Thus,
the current-assisted Raman activation can successfully explain this
experiment and an extension of the setup
allows to demonstrate the equivalence of the dynamically induced dc component
and an external dc current.

\begin{figure}[t]
    \centering
    \includegraphics[width=\columnwidth]{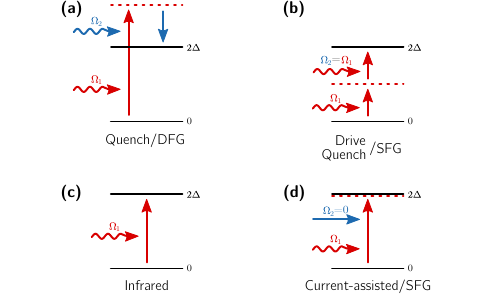}
    \caption{\label{fig:excitation}%
    Different excitation schemes of Higgs mode.
    a) Excitation due to quench with two-photon Raman
    difference-frequency generation (DFG) process.
    The frequencies $\Omega_1$ and $\Omega_2$
    are within the bandwidth of the quench pulse.
    b) Driving of Higgs mode with two-photon Raman
    sum-frequency generation (SFG) process.
    The frequencies of the photons are equal $\Omega_1 = \Omega_2 = \Omega$.
    Alternatively, a quench can be implemented as a SFG process as well,
    where the two frequencies are within the bandwidth of the quench pulse.
    c) One-photon infrared process.
    d) Two-photon current-assisted Raman process
    with $\Omega_1 = \Omega$ and $\Omega_2 = 0$
    reflecting a special case of a SFG process.
    }
\end{figure}

\textit{SFG and DFG for Higgs mode.}
\label{sec:sdfg}
With this insight, we can summarize and classify
the possible known excitation schemes of Higgs modes
as sketched in Fig.~\ref{fig:excitation}.
In the impulsive excitation of a Higgs mode,
a short intense THz pulse quenches the Mexican hat potential
of the complex superconducting order parameter.
That process follows the scheme of impulsive stimulated Raman scattering
\cite{JChemPhys.83.5391,PhysRevB.65.144304}
shown in Fig.~\ref{fig:excitation}(a).
The Higgs mode is excited via the difference-frequency
of the photons $\Omega_1$ and $\Omega_2$ that stem from the same pulse.
The required frequencies are within the bandwidth
of the ultrashort broadband THz pulse.
Experimentally, the free Higgs oscillations observed in NbN
\cite{PhysRevLett.111.057002} are excited this way.

Instead of a difference-frequency process between the incoming photons,
it is also possible to excite the Higgs mode via a sum-frequency process
to excite Raman active modes \cite{PhysRevLett.119.127402}
as shown in Fig.~\ref{fig:excitation}(b).
In analogy to Fig.~\ref{fig:excitation}(a),
this can be a quench of the system,
where the two photons stem from the same pulse.
Furthermore, the $2\Omega$ oscillations of the driven Higgs mode in NbN
\cite{Science.345.1145},
Nb$_3$Sn \cite{NatPhoton.10.707,PhysRevLett.124.207003}
and in cuprates \cite{PhysRevLett.120.117001,NatCommun.11.1793},
as well as the driven Leggett mode in MgB$_2$ \cite{NatPhys.15.341}
can be also understood in this way.
In these cases, the two photons $\Omega_1 = \Omega_2 = \Omega$
have the same frequency
leading to the experimentally observed second harmonic generation
and the $2\Omega=2\Delta$ resonance condition.
This two-photon Raman process is described
by the $\propto A_iA_j$ term in Eq.~\eqref{eq:eps_expansion}.

In contrary, an infrared excitation is a one photon absorption process
by coupling to a dipolar moment shown in Fig.~\ref{fig:excitation}(c).
However, as we have discussed in this communication,
the current-driven superconductor does not change
the character of the Higgs mode.
The current rather gives rise to a new $\propto A_iQ_j$ term
in Eq.~\eqref{eq:eps_expansion}
that describe an effective Raman two-photon excitation.
In addition to the photon $\Omega_1=\Omega$,
the current takes the role of a photon with $\Omega_2=0$.
As such, the current-driven case can be understood in analogy
with the SFG or DFG Raman processes,
leading to $\Omega_1 \pm \Omega_2 = \Omega$
and the new $\Omega=2\Delta$ resonance condition
shown in Fig.~\ref{fig:excitation}(d).

\textit{Conclusion.}
In short, we provide an alternative explanation
for the recent experimental conclusion
that in the presence of a supercurrent
the Higgs mode becomes infrared active.
Our theory shows that in this case, the activation of the Higgs mode
is an effective Raman process (SFG or DFG),
where one of the photons is a virtual photon at $\omega = 0$.
On the other hand we demonstrate that an infrared activation is negligible.

Our theory properly describes the experimental observation
of the appearance of the Higgs mode in the linear THz spectrum of NbN
in the presence of a dc current \cite{PhysRevLett.122.257001}.
Moreover, it explains the appearance of the $\Omega$ oscillation
in the THz driven Nb$_3$Sn in addition to the $2\Omega$ terms
and the higher odd order interference terms
\cite{NatPhoton.10.707,PhysRevLett.124.207003},
resulting from a dynamically driven dc current.
With a model calculation
we propose an extension to this experiment
which allows to confirm the equivalence
of a dynamically driven dc supercurrent and our theory
by a suppression of odd order higher harmonics in the gap oscillations.
Furthermore,
our results are not restricted to conventional $s$-wave superconductors
and thus, will guide further current-assisted experiments
also on unconventional superconductors.

\textit{Acknowledgements.}
We thank the Max Planck-UBC-UTokyo Center
for Quantum Materials for fruitful collaborations and
financial support.

\end{document}